\documentclass[twoside,12pt]{article}
\usepackage{epsfig}

\newcommand{\be}{\begin{equation}}
\newcommand{\ee}{\end{equation}}
\newcommand{\bea}{\begin{eqnarray}}
\newcommand{\eea}{\end{eqnarray}}

\begin{document}

\begin{titlepage}

\begin{center}
{\large \bf The dependence of information entropy of uniform Fermi
systems on correlations and thermal effects}

\vspace{9mm}

{Ch.C. Moustakidis and S.E. Massen} \\
\vspace{5mm}
{\it Department of Theoretical Physics, Aristotle University of Thessaloniki,
GR-54124 Thessaloniki, Greece }


\end{center}


\vspace{7mm}


\begin{abstract}
The influence of correlations of uniform Fermi systems (nuclear
matter, electron gas and liquid $^3$He) on Shannon's information
entropy, $S$, is studied. $S$ is the sum of the information
entropies in position and momentum spaces. It is found that, for
three different Fermi systems with different particle
interactions, the correlated part of $S$ ($S_{cor}$) depends on
the correlation parameter of the systems or on the discontinuity
gap of the momentum distribution through two parameter
expressions. The values of the parameters characterize the
strength of the correlations. A two parameter expression also holds
between $S_{cor}$ and the mean kinetic energy ($K$) of the Fermi
system. The study of thermal effects on the uncorrelated electron
gas leads to a relation between the thermal part of $S$
($S_{thermal}$) and the fundamental quantities of temperature,
thermodynamical entropy and the mean kinetic energy.
It is found that, in the case of low temperature limit,
the expression connecting $S_{thermal}$ with $K$ is the same to the one
which connects $S_{cor}$ with $K$.
There are only
some small differences on the values of the parameters. Thus,
regardless of the reason (correlations or thermal) that changes
$K$, $S$ takes almost the same value.

\vspace{5mm}


\end{abstract}
\end{titlepage}

\section{Introduction}
Information theoretical methods have in recent years played an
important role in the study of quantum mechanical systems
\cite{Bialy75,Gadre84,Ohya93,%
BAB95,Nagy96,Majer96,Panos97,Massen98} in two cases: first in the
clarification of fundamental concepts of quantum mechanics and
second in the synthesis of probability densities in position and
momentum space. An important step was the discovery of an entropic
uncertainty relation \cite{Bialy75} which for a three-dimensional
system has the form
\begin{equation}
S= S_r + S_k \ge 3 (1+ \ln \pi) \simeq 6.434,
 \label{S1}
\end{equation}
where
\begin{equation}
S_{r}=-{\int}{\rho}({\bf r}) \ln {\rho}( {\bf r})d {\bf r},
\qquad  S_{k}=-{\int}n( {\bf k})\ln n({\bf k})d {\bf k} \label{4}
\label{Sr-1}
\end{equation}
%
are Shannon's information entropies (IE) in position- and
momentum-space and $\rho ({\bf r})$, $n({\bf k})$ are the density
distribution (DD) and momentum distribution (MD), respectively,
normalized to unity.

The physical meaning of $S_r$ and $S_k$ is that it is a measure of
quantum-mechanical uncertainty and represents the information
content of a probability distribution, in our case of various
fermionic systems density and momentum distributions. Inequality
(\ref{S1}) provides a lower bound for  $S$ which is attained for
Gaussian wave functions. It is mentioned that the sum $S=S_r+S_k$
is invariant to uniform scaling of coordinates, while the
individual entropies $S_r$ and $S_k$ are not.

Ziesche \cite{Ziesche95}, mentions in his paper that March refers
to the information entropy with the following words: ``Further
work is called for before the importance of $S_r$ and $S_k$ in
atomic theory can be assessed'' \cite{March}. We could extend that
statement for fermionic and correlated bosonic systems as well.

The motivation of the present work, which is in the spirit of the
above statement, is to extend our previous study of  IE in nuclei,
atomic clusters and correlated bosonic systems to the direction of
various uniform fermionic systems and to connect it with the
interaction of the particles and the temperature. In uniform
systems the density $\rho=N/V$ is a constant and the interaction
of the particles is reflected to MD which deviates from the
$theta$ function form of the ideal Fermi-gas model. It is
important to study how the interaction affects the MD as well as
the IE. An attempt is also made to relate the IE with  fundamental
quantities such as the temperature, the thermodynamical entropy
and the mean kinetic energy of the fermionic system (electron
gas).

The quantum systems which are examined in the present work are
nuclear matter, electron gas and liquid $^3$He.
Traditionally, the helium system is regarded as the most
strongly interacting, with an almost-hard-core interaction, and
the electron gas as the most weakly interacting, the inner Coulomb
repulsion being very soft by comparison. The nuclear case lies
somewhere  between. Historically, the helium problem has proved
to be the hardest to be solved, because of the hard core interaction. On
the other hand, the nuclear problem presents a singular
frustration and a special challenge because of the strong state
dependence (dependence on spin, isospin, angular momentum) and
noncentral character of the bare two-nucleon interaction, and
also because of ambiguities in the determination of the
interaction from first principles \cite{Silver89}.
Moreover, the judgment about
strong versus weak interaction depends on the density being studied. This
fact is vividly illustrated by the electron gas, which is
distinguished from the other examples by the long-range nature of
the Coulomb interaction. Consequently, strong coupling prevails in
the electron gas in the limit of low density, whereas  the helium and
the nuclear systems become more strongly interacting as the density
increases. In all these cases the strength of the interaction may
be gauged by the depletion of the Fermi sea. Quantitatively, one
examines the deviation of $Z_F$ from unity, where $Z_F$ is
the discontinuity gap of the momentum distribution $n(k)$ at $k=k_F$ in
an uniform Fermi system.

The paper is organized as follows. The method leading to the
expression of Shannon's information entropy sum in finite Fermi
systems is presented in Section II. Applications of that
expression to nuclear matter, electron gas, and liquid $^3$He are
made in the three subsections of Section II.  In the same
subsections  numerical results are also reported and discussed.
In Section III the study of the influence of thermal effects on
the information entropy sum is made. Finally, the concluding
remarks and the Summary of the present work are given in Section
IV.

\section{Information entropy for an infinite Fermi system}
The key quantity for the description of the MD
both in infinite and finite quantum systems is the one-body
density matrix (OBDM). The OBDM is defined as
\begin{equation}
\rho({\bf r}_1,{\bf r}_1')=
\int \Psi^{*}({\bf r}_1,{\bf r}_2,...,{\bf r}_N)
\Psi({\bf r}_1',{\bf r}_2,...,{\bf r}_N)
d{\bf r}_2 ... d{\bf r}_N
\end{equation}
The diagonal elements $\rho({\bf r_1},{\bf r_1})$ of the OBDM
yields the local density distribution, which is
just a constant $\rho$ in the uniform infinite system. Homogenity and
isotropy of the system require that $\rho({\bf r}_1,{\bf r}_1')=
\rho(\mid {\bf r}_1-{\bf r}_1'\mid)\equiv \rho(r)$. In the case of
noninteractive Fermi systems the associated OBDM is
\[\rho(r)= \rho l(k_F\mid{\bf r}_1-{\bf r}_1'\mid), \]
where
\[
l(x)=3x^{-3}(\sin x-x \cos x) \]
and $\rho=N/V$ is the constant density of the uniform Fermi system.

The density, normalized to $1$ ($\int \rho_0 d {\bf r}=1$), is given
by the relation
\begin{equation}
\rho_o=\frac{1}{N V_o}=\frac{1}{N \frac{4}{3}\pi r_o^3}
\end{equation}
where the volume $V_o=\frac{4}{3}\pi r_o^3$ corresponds to the
effective volume of the Fermi particle and $N$ is the number
of fermions.

The MD for fermions, having single-particle level
degeneracy $\nu$, is defined by
\begin{equation}
n(k)=\nu^{-1} \int \rho(r) e^{i {\bf kr}} d{\bf r}
\end{equation}
The MD, normalized to $1$ ($\int n(k) d {\bf k}=1$),
is given by the relation
\begin{equation}
n(k)=\frac{1}{V_k}\left\{ \begin{array}{cc}
    \tilde{n}(k^{-})   , &  \mbox{$k<k_F$ } \\
     \tilde{n}(k^{+})      , &  \mbox{$k>k_F$}
                              \end{array}
                       \right.
\label{Cor-MD}
\end{equation}
where $V_k=\frac{4}{3}\pi k_F^3$. The Fermi wave number $k_F$ is
related with the constant density $\rho=N \rho_0=3/(4\pi r_0^3)$
as follows
\begin{equation}
k_F=\left( \frac{6 \pi^2 \rho}{\nu} \right)^{1/3}=
\left(\frac{9 \pi}{2 \nu}\frac{1}{r_0^3} \right)^{1/3}
\label{kfermi}
\end{equation}
where $\nu=2$ for electron gas and liquid $^3$He and  $\nu=4$ for
nuclear matter. In the case of an ideal Fermi gas the MD has the
form
\begin{equation}
n_0(k)=\frac{1}{V_k}\theta(k_F-k)
\label{MD-IG}
\end{equation}

The information entropy in coordinate space (for density $\rho_0$
normalized to $1$)
for a correlated or uncorrelated Fermi system is given by the relation
\begin{equation}
S_r=-\int \rho_o \ln \rho_o d{\bf r}=\ln V .
\label{IE-Sr-1}
\end{equation}
Considering that $V=N V_o$, $S_r$ becomes
\begin{equation}
S_r=\ln \frac{4}{3}\pi r_o^3 +\ln N .
\label{IE-Sr-2}
\end{equation}

The information entropy in momentum space (for n(k) normalized
to 1) is given by the relation
\begin{equation}
S_k=-\int n(k) \ln n(k) d{\bf k} .
\label{IE-Sk-1}
\end{equation}
$S_k$ for an ideal Fermi gas, using Eq. (\ref{MD-IG}), becomes
\begin{equation}
S_k=\ln V_k=\ln \left(\frac{6 \pi^2}{\nu}\frac{1}{r_0^3} \right)
\label{IE-Sk-2}
\end{equation}
From Eq. (\ref{IE-Sr-2}) and (\ref{IE-Sk-2}) the information
entropy sum $S=S_r+S_k$ for an uncorrelated infinite Fermi system becomes
\begin{equation}
S_0=S_r+S_k=\ln \left(\frac{8 \pi^3}{\nu} \right)+\ln N
\label{IE-S0}
\end{equation}

It turns out that the functional form
\[
S_0=a+b \ln N
\]
for the entropy sum as a function of the number of particles
$N$ holds for the ideal infinite Fermi systems. The same function has been
found in Ref. \cite{Gadre84} for atoms and in Ref. \cite{Massen98} for nuclei
and atomic clusters. That expression has been found also
in Ref. \cite{Panos03} for the ideal electron gas.

In the case of correlated Fermi systems, the IE in coordinate space
is given again by Eq. (\ref{IE-Sr-2}) while the IE in momentum space
can be found from Eq. (\ref{IE-Sk-1}) replacing
$n(k)$ from  Eq. (\ref{Cor-MD}). $S_k$ is written now
\begin{equation}
S_k=\ln V_k-\frac{4 \pi}{V_k}\left(
\int_{0}^{k_F^{-}} k^2 \tilde{n}(k^{-})\ln\tilde{n}(k^{-})  dk+
\int_{k_F^{+}}^{\infty} k^2 \tilde{n}(k^{+}) \ln\tilde{n}(k^{+})  dk \right).
\label{Cor-Sk-1}
\end{equation}
The correlated entropy sum has the form
\begin{equation}
S=S_r+S_k=S_0+S_{cor}
\label{Cor-S-1}
\end{equation}
where $S_0$ is the uncorrelated entropy sum of Eq. (\ref{IE-S0})
and $S_{cor}$ is the contribution of the particles correlations to
the entropy sum. That contribution can be found from the expression
\begin{equation}
S_{cor}=-3\left(
\int_{0}^{1^{-}} x^2 \tilde{n}(x^{-})\ln\tilde{n}(x^{-}) dx+
\int_{1^{+}}^{\infty} x^2 \tilde{n}(x^{+})\ln\tilde{n}(x^{+}) dx \right),
\label{S-cor}
\end{equation}
where $x=k/k_F$.

Another quantity expected to be related with the IE is the mean
kinetic energy $K$, defined by
\begin{eqnarray}
K&=&\frac{\hbar^2}{2m} \int n(k) k^2 d{\bf k}
=3 \epsilon_F \int_{0}^{\infty}x^4 \tilde{n}(x) d x
\nonumber\\
&=&3
\epsilon_F \left(\int_{0}^{1^{-}} x^4 \tilde{n}(x^{-}) dx+
\int_{1^{+}}^{\infty} x^4 \tilde{n}(x^{+}) dx \right)
\label{KE-1}
\end{eqnarray}
where $\epsilon_F=\hbar ^2 k_F^2/(2m)$
is the Fermi energy.

From the above analysis it is clear that in order to calculate the
IE sum in  uniform Fermi systems, the knowledge of the MD is required.
In dealing with various fermionic systems, we are necessarily driven to
computational methods, since a pure analytical treatment is
intractable. Computational many-body methods may be classified in
several different ways: they may be  based on wave functions or
on field theory. They may be variational or perturbative. In any
case the singular, or near-singular character of the basic forces
involved precludes a simple, stepwise perturbative calculation
within an independent-particle (plane-wave) basis.

In the present work  we apply the low order approximation (LOA) for the
calculation of the MD in nuclear matter
\cite{Gaudin71,Dalri82,Flyn84}. For liquid  $^3$He we use the
results of Moroni et al.
\cite{Moroni97}, while the MD for the electron gas is taken from a work
of P. Gori-Giorgi et al. \cite{Paola-02}.
Of course there are  a lot of
data for the MD, for the two cases, in the literature. In our work we
try to use the modern ones  that exist up to now. It is
worthwhile to point out also that our primary purpose is not the
detailed analysis of the MD but the accurate calculation of the
correlated part of the
information entropy in various cases, using reliable
data for the MD.


\subsection{Nuclear matter}

The model we study is based on the Jastrow ansatz for
the ground state wave function of nuclear matter
\begin{equation}
\Psi({\bf r}_1,{\bf r}_2,...,{\bf r}_N)=
\prod_{1 \leq i \leq  j  \leq N} f(r_{ij})
\Phi({\bf r}_1,{\bf r}_2,...,{\bf r}_N)
\label{Jastrow-1}
\end{equation}
where $r_{ij}=\mid {\bf r}_i-{\bf r}_j \mid$, $\Phi$
is a Slater determinant (here, of plane waves with appropriate
spin-isospin factors, filling the Fermi sea) and $f(r)$ is  a
state-independent two-body correlation function.
In the present work the correlation function is taken to be the Jastrow
function \cite{Jastrow55}
\begin{equation}
f(r)=1-\exp[-\beta^2 r^2]
\label{beta-Gaus}
\end{equation}
where $\beta$ is the correlation parameter.
A cluster expansion for the one-body density matrix
$\rho({\bf r}_1,{\bf r}_1')$ has been derived by
Gaudin, Gillespie and Ripka \cite{Gaudin71,Dalri82,Flyn84} for the Jastrow
trial function (\ref{Jastrow-1}).

In the LOA the momentum distribution is constructed as \cite{Flyn84}
\begin{equation}
n_{LOA}(k)=\theta(k_F-k) \left[ 1-k_{dir}+Y(k,8) \right]+
8 \left[ k_{dir}Y(k,2)-[Y(k,4)]^2 \right]
\label{mn-mom-1}
\end{equation}
where
\begin{equation}
c_{\mu}^{-1}Y(k,\mu)=
\frac{e^{-\tilde{k}_{+}^{2}}-e^{-\tilde{k}_{-}^{2}}}{2\tilde{k}}
+\int_{0}^{\tilde{k}_{+}} e^{-y^2} dy+
{\rm sgn}(\tilde{k}_{-}) \int_{0}^{\mid \tilde{k}_{-} \mid} e^{-y^2} dy
\end{equation}
and
\begin{equation}
c_{\mu}=\frac{1}{8\sqrt{\pi}}\left(\frac{\mu}{2}\right)^{3/2},
\quad
\tilde{k}=\frac{k}{\beta \sqrt{\mu}} ,
\quad
\tilde{k}_{\pm}=\frac{k_{F}\pm k}{\beta \sqrt{\mu}},
\quad
\mu=2,4,8.
\end{equation}
and ${\rm sgn}(x)=x/\mid x \mid$.
The normalization condition for the momentum distribution is
\begin{equation}
\int_{0}^{\infty} n_{LOA}(k)k^2 dk=\frac{1}{3}k_{F}^{3}
\end{equation}

A rough measure of correlations and of the rate of convergence of the
cluster expansion is given by the dimensionless Jastrow wound
parameter
\begin{equation}
k_{dir}=\rho \int [f(r)-1]^2 d {\bf r}
\label{eq-kdir}
\end{equation}
where $\rho=2k_{F}^{3}/(3\pi^2)$ is the density of the uniform
nucleon matter. Eq. (\ref{eq-kdir}) gives the following relation
between the wound parameter $k_{dir}$ and the correlation parameter
$\beta$
\begin{equation}
k_{dir}=\frac{1}{3 \sqrt{2\pi}}\left(\frac{k_F}{\beta} \right)^3
\end{equation}
It is clear that large $k_{dir}$ implies strong correlations and
poor convergence of the  cluster expansion. In the numerical
calculations the correlation parameter $\beta$ was in the
interval: $1.01 \leq \beta \leq  2.482 $. That range corresponds
to $ 0.3 \geq k_{dir}  \geq 0.02 $ and is a reasonable
interval in the case of nuclear matter \cite{Flyn84}.

The calculated values of $S_{cor}$ for nuclear matter versus the
wound parameter $k_{dir}$ are displayed by points in Fig.1a. It is
seen that $S_{cor}$ is an increasing function of $k_{dir}$. The
function $S_{cor}(k_{dir})$ is equal to zero for $k_{dir}=0$ (no
correlations) and the dependence of $S_{cor}$ on $k_{dir}$ is not
very far from a linear dependence. Thus we fitted the
numerical values of $S_{cor}$ with the two parameters formula
\begin{equation}
S_{cor}(k_{dir})=s k_{dir}^{\lambda}
\end{equation}
That simple formula, with the best fit values of the parameters
\[s=2.0575, \qquad \lambda=0.6364 \]
reproduces the numerical values of $S_{cor}$ very well.

Another characteristic quantity which is used as a measure of the
strength of correlations of the uniform Fermi systems is the
discontinuity, $Z_F$, of the MD at $k/k_F=1$. It is defined as
\[ Z_{F}=n(1^{-})-n(1^{+}) . \]

For ideal Fermi systems $Z_F=1$, while for interacting ones
$Z_F<1$. In the limit of very strong interaction $Z_F=0$
there is no discontinuity on the MD of the system. The quantity
($1-Z_F$) measures the ability of correlations to deplete the Fermi
sea by exciting particles from states below it (hole states) to
states above it (particle states) \cite{Flyn84}.

The dependence of $S_{cor}$ on the quantity $(1-Z_F)$ is shown in
Fig. 2. It is seen that $S_{cor}$ is an increasing function of
$(1-Z_F)$. For the same reasons mentioned before we fitted the
numerical values of $S_{cor}$ to the two parameters formula
\begin{equation}
S_{cor}(Z_F)=s(1-Z_F)^{\lambda}
\label{Scor-Zf-NM}
\end{equation}
As before, the above simple formula, with the best fit values
of the parameters
\[s=2.2766, \qquad \lambda=0.6164 \]
reproduces the numerical values of $S_{cor}$ very well.

From the above analysis we can conclude that the correlated part
of the information entropy sum can be used as a measure of the
strength of correlations in the same way the wound parameter
and the discontinuity parameter are used.

An explanation of the above behaviour of $S_{cor}$ is the
following: The effect of nucleon correlations is the departure
from the step function form of the MD (ideal Fermi gas) to the one
with long tail behaviour for $k>k_F$. The diffusion of the MD
leads to a decrease of the order of the system (in comparison to the
ordered step function MD), thus it leads to an increase of the
information content of the system.

Concluding we should state that, the increase of information
entropy sum of the nuclear matter is due to the increase of the
number of nucleons of the system, as it is seen from
Eq. (\ref{IE-S0}) and also to the increase of correlations.

Finally, the dependence of the IE on the kinetic energy $K$, which
is given by Eq. (\ref{KE-1}), is also examined. The calculated
values of the correlated part of IE, $S_{cor}$, versus $K$ is shown
in Fig. 3. $S_{cor}$ is an increasing function of $K$. It should
be noted that $S_{cor}$ is equal to $0$ for
$K=\frac{3}{5}\epsilon_F$. For that reason we fitted the numerical
values of $S_{cor}$ to the formula
\begin{equation}
S_{cor}(K)=s \left( \frac{K}{\epsilon_F}-0.6 \right)^{\lambda}.
\label{Scor-KE-NM}
\end{equation}
That simple formula, with the best fit values of the parameters
\[s=3.7413, \qquad \lambda=1.5911 \]
reproduces the numerical values of $S_{cor}$ very well.

Summarizing, we can conclude that $S_{cor}$ in  nuclear matter
is an increasing function of the wound parameter and the discontinuity
parameter ($1-Z_F$). It is also an increasing function of
the mean kinetic energy
of the system. The dependence of $S_{cor}$ on those quantities is
given by simple two parameter formulae.

\subsection{Electron gas}

We consider the electron gas as a system of fermions interacting
via a Coulomb potential. The electron gas is a model of the
conduction electrons in a metal where the periodic positive potential
due to the ions is replaced by a uniform charge distribution.
The density of the uniform electron gas (Jellium) is
$\rho=3/(4\pi r_o^3)$ and the momentum distribution is $n(x,r_s)$,
where $x=k/k_F$ and $r_s=r_o/a_B$ (with $a_B=\hbar^2/me^2$, the
Bohr radius).

The momentum distribution of the unpolarized uniform electron gas
in its Fermi-liquid regime, $n(x,r_s)$ is constructed with the help
of the convex Kulik function $G(\chi)$ \cite{Paola-02}. It is
assumed that $n(0,r_s)$, $n(1^{\pm},r_s)$, the on-top pair density
$g(0,r_s)$, and the kinetic energy $K(r_s)$ are known
(respectively, from accurate calculations for $r_s=1,\ldots,5$,
from the solution of the Overhauser model, and from quantum Monte
Carlo calculations via the virial theorem) \cite{Paola-02}.

The qualitative behavior of $n(x,r_s)$ is the following. It starts
at $x=0$ with a value $n(0,r_s) \leq 1$, and decreases with
increasing $x$. For $x<1$, it is concave. Then in the Fermi liquid
regime at $x=1$, there is a finite jump (Fermi gap) from
$n(1^{-},r_s)$ to a lower value
$n(1^{+},r_s)=n(1^{-},r_s)-Z_F(r_s)$ with logarithmic slopes at
both sides of $x=1$. For $x>1$, (correlation tail) $n(x,r_s)$ is
convex and vanishes for $x \rightarrow \infty$. For $r_s=0$ (ideal
Fermi gas), $n(x,r_s)$ has the well known step function form
$n(x,0)=\theta(1-x)$. Thus, the discontinuity $Z_F(r_s)$, starts
with $Z_F(0)=1$ and decreases with increasing interaction strength
$r_s$. The discontinuity $Z_F$ of $n(k)$ at the Fermi surface
narrows as the density decreases, which implies that the system is
becoming more strongly coupled. That behavior is due to the fact
that the screening of the long-range Coulomb interaction between
the electrons becomes less effective at lower density. The
inverse behaviour appears in nuclear matter cases and the
atomic $^3$He, where the basic interactions are of short range
and $Z_{F}$ decreases as the density increases.

At large  $r_s$, the electrons
form a Wigner crystal with a smooth $n(x,r_s)$. $r_s \ll 1$ and $r_s \gg 1$
are the weak- and strong-correlation limits, respectively. For intermediate
values of $r_s$, a non-Fermi liquid regime may exist with $Z_F=0$.
In such a case, $n(x,r_s)$ would be continuous versus $x$,
with a nonanalytical
behavior at $x=1$ \cite{Paola-02}.

We examined the dependence of the correlated part of the IE for the electron
gas on the correlation parameter $r_s$, (or $\rho=3/(4\pi r_o^3)$),
the discontinuity parameter ($1-Z_F$) and the mean kinetic energy $K$.
The dependence of $S_{cor}$ on those parameters are shown in Fig. 1b, 2, 3.
It is seen that, as in the case of nuclear matter,
$S_{cor}$ depends on those quantities through two
parameter expressions
of the form
\begin{equation}
S_{cor}(r_s)=s r_s^{\lambda}
\label{Scor-ro-EG}
\end{equation}
with
\[s=0.1312, \qquad \lambda=0.8648 ,\]

\begin{equation}
S_{cor}(Z_F)=s(1-Z_F)^{\lambda}
\label{Scor-ZF-EG}
\end{equation}
with
\[s=2.0381, \qquad \lambda=1.6899\]
and
\begin{equation}
S_{cor}(K)=s \left(\frac{K}{\epsilon_F}-0.6 \right)^{\lambda}
\label{Scor-KE-EG}
\end{equation}
with
\[s=2.0786, \qquad \lambda=0.6601.\]

The values of the parameters $s$ and $\lambda$ have been found by
least squares fit of the above expressions to the calculated values
of $S_{cor}$.

\subsection{Liquid  $^3$He}

The helium interaction potential is very strong at  small
distances, its core repulsion being very hard (but not infinite).
As a consequence there is a Fermi-surface discontinuity of roughly
$Z_F\sim 0.3$. This small value supports the view that liquid
$^3$He is the most strongly interacting  Fermi system we have
considered.

In the case of liquid $^3$He the calculation of the
momentum distribution is performed from diffusion Monte Carlo
(DMC) simulations using trial functions, optimized via the Euler
Monte Carlo (EMC) method \cite{Moroni97}. The EMC wave functions
have pair and triplet correlations fully optimized, and provide
the lowest available energy bounds. Moreover, their use in DMC
calculations has led to results of unprecedented accuracy for the
energy, pair function, and static structure function. For $^3$He,
backflow correlations have been included, in the usual way, by
replacing the plane waves $\exp(i{\bf k}_i{\bf r}_j) $ in the
Slater determinant with $\exp(i{\bf k}_i{\bf \chi}_j) $, where
${\bf \chi}_j={\bf r}_j+\sum_{k \neq j} \eta (r_{ij}({\bf r}_j-{\bf
r}_k))$. The function $\eta(r)$ can be taken either short range, or
long range. The $^3$He results, presented below, were obtained with
short-range backflow.

As in the cases of nuclear matter and electron gas, we examined the
dependence of the correlated part of the IE for the liquid $^3$He
on the density $\rho=3/(4\pi r_o^3)$, the discontinuity parameter
$(1-Z_F)$ and the mean kinetic energy $K$. The dependence of those
parameters are shown in Figures 1c, 2 and 3, respectively. As in
the previous two cases, $S_{cor}$ depends on those parameters
through simple two parameter formulae of the form
\begin{equation}
S_{cor}(\rho)=s \rho^{\lambda}
\label{Scor-rho-HL}
\end{equation}
with
\[s=2032.56, \qquad \lambda=1.4757 \]

\begin{equation}
S_{cor}(Z_F)=s(1-Z_F^{\lambda})
\label{Scor-Zf-HL}
\end{equation}
with
\[s=13.0640, \qquad \lambda=0.2070\]
and
\begin{equation}
S_{cor}(K)=s \left(\frac{K}{\epsilon_F}-0.6 \right)^{\lambda}
\label{Scor-KE-HL}
\end{equation}
with
\[s=3.0993, \qquad \lambda=0.5236 .\]
The values of the parameters $s$ and $\lambda$ have been found by
least squares fit of the above expressions to the calculated values of
$S_{cor}$. The values of the parameters of expressions
(\ref{Scor-rho-HL}) and (\ref{Scor-Zf-HL}) indicate the strong character
of the interaction of liquid $^3$He. That character is also indicated by
the expression of $S_{cor}(Z_F)$ (Eq. (\ref{Scor-Zf-HL})). That expression
differs from the corresponding expressions of the electron gas and
nuclear matter.

\section{Thermal effects in electron gas}
The electrons of the electron gas, at temperature $T=0$, occupy
all the lower available states up to a highest one, the Fermi level.
As the temperature increases the electrons of the gas tend to become excited
into states of energy of order $kT$ higher than the Fermi energy.
However, the electrons
with the lower energy cannot be excited as there are not available states
for them to be excited. Only a small fraction of the gas, of order
$T/T_F$, with energy about $kT$ lower than the Fermi energy have
any chance to be excited. The rest remain unaffected in their
zero-degree situation.
The net result is that the mean occupation
number becomes slightly blurred compared to its sharp, step
function form at $T=0$ \cite{Goodstein}. In general the occupation
number of the  electron gas is given by the Fermi-Dirac formula
\begin{equation}
n(\epsilon)=\frac{1}{\exp\left[\frac{1}{k_B T}(\epsilon-\mu) \right] +1}
\label{Fermi-Dirac}
\end{equation}
where $\epsilon=\frac{p^2}{2m}$ ($p=\hbar k$), $k_B$ is the
Boltzmann's constant and $\mu$ is the chemical
potential. The chemical potential of a gas at absolute zero
($T=0$) coincides with the Fermi energy $\epsilon_F$. This is the
characteristic energy for a Fermi gas and is by definition
the energy of the highest single-particle level occupied at $T=0$.
The Fermi energy is given by the relation
\begin{equation}
\epsilon_F=\frac{\hbar^2}{2m}\left(3\pi^2 \rho \right)^{2/3}
\end{equation}
while the Fermi temperature is defined by
\begin{equation}
\epsilon_F=k_B T_F
\end{equation}
We will examine how the IE sum of the electron gas is affected when
the temperature starts to increase above zero.
Our study will include the cases of
low temperature and  high temperature limit, separately.

\subsection{Thermal effects in electron gas for $T\ll T_F$ }

Since there is only one characteristic temperature, the Fermi
temperature, by the term low energy we will mean the limit $T\ll T_F$.
It is easy to see that for electron  gas, i.e. in copper,
$T_F \sim 8.5 \times 10^4$ $^{o}K$, while the melting point
is of the order of $10^3$ $ ^{o}K$. Thus, at all temperatures at
which copper is a solid, the condition $T\ll T_F$ is satisfied;
the electron gas is in its low-temperature limit.
For $T\ll T_F$ the chemical potential, in a first approximation, is
\cite{Goodstein,Mandl,Huang}
\begin{equation}
\mu=\epsilon_F\left[1-\frac{\pi^2}{12}\left(\frac{T}{T_F}\right)^2 \right]
\end{equation}
and so Eq. (\ref{Fermi-Dirac}) becomes
\begin{equation}
n(x)=\frac{1}{\exp\left[\frac{1}{\xi}
\left(x^2-1+\frac{\pi^2}{12}\xi^2\right)\right]+1}
\label{nx-1}
\end{equation}
where $x=(\epsilon/\epsilon_F)^{1/2}=k/k_F$ and $\xi=T/T_F\ll 1$.
The normalization of  $n(x)$ is
$\int_{0}^{\infty} x^2n(x) dx=1/3$.

Following the same steps as in Section 2, the information entropy sum
of the electron gas at temperature $T\ll T_F$ is written
\begin{equation}
S=S_0+S_{thermal}
\end{equation}
where $S_0$ is given by Eq.(\ref{IE-S0}) and
\begin{equation}
S_{thermal}=-3
\int_{0}^{\infty} x^2 n(x) \ln n(x)  dx
\end{equation}

It is worthwhile to notice that the correlations between the fermi
particles invoke discontinuity to the MD at $k=k_F$ while the
thermal effect causes just a slight deviation from the sharp
step function form at $T=0$. That is shown in Fig. 4a, where the MD for
a correlated electron gas with $r_s=5$ and for an ideal electron gas at
temperature $T=0$ and $T/T_F=0.2$  have been plotted versus $k/k_F$. The
two cases of the figure ($r_s=5$ and $T/T_F=0.2$) give the same value
for the information entropy. Thus, even though the origin of the two
effects (correlations and temperature) is different and they influence
in a different way the MD, the two information entropies
$S_{cor}$ and $S_{thermal}$ are almost the same.

The calculated values of $S_{thermal}$ for various values of the
temperature ( for $T\ll T_F$ ) are shown in Fig. 4b. It is seen that
$S_{thermal}$ is an increasing function of the temperature and
depends linearly on it. The line
\begin{equation}
S_{thermal}=\alpha \left(\frac{T}{T_F} \right), \qquad \alpha=2.5466
\label{IE-FG}
\end{equation}
reproduces very well all the calculated values of $S_{thermal}$.
That expression of the information entropy is similar to the expression which
gives the thermodynamical entropy, $S_{TE}$, for $T\ll T_F$. $S_{TE}$
in the low temperature limit
has the form \cite{Goodstein,Huang}
\begin{equation}
S_{TE}=\frac{\pi^2}{2}Nk_B\frac{T}{T_F}
\label{TE-FG}
\end{equation}
Comparing  Eqs. (\ref{IE-FG}) and (\ref{TE-FG}), a relation between
the two entropies could be found in the case $T\ll T_F$.
That relation has the form
\begin{equation}
S_{thermal}=\frac{2 \alpha}{\pi^2} \frac{S_{TE}}{N k_B},
\qquad \alpha=2.5466
\label{IE-TE-FG-1}
\end{equation}
while the information entropy sum is written
\begin{equation}
S_{IE}=\ln N + \ln 4 \pi^3 + \frac{2 \alpha}{\pi^2} \frac{S_{TE}}{N k_B}
\label{IE-TE-FG-2}
\end{equation}
Thus, the information entropy of a Fermi gas, which is a measure
of the information content of the system, depends on the number of fermions
as well as on the thermodynamical entropy of the system.

The increase of the temperature changes also the mean
kinetic energy $K$ of the ideal electron gas. For $T\ll T_F$,
$K$ is given by \cite{Huang}
\begin{equation}
K=\frac{3}{5}\epsilon_F \left[1+
      \frac{5}{12}\pi^2\left(\frac{T}{T_F}\right)^2\right]
\label{K-LTL}
\end{equation}
In the examined range of $T$, $K$ changes about $15$ \%.
As $K$ appears both in correlated and uncorrelated Fermi systems, and
a relation of the form $S_{cor}=S_{cor}(K)$ was already found in Sec. 2,
it is of interest to examine the existence of a relation between
$S_{thermal}$ and $K$.

That relation can be easily found writing Eq. (\ref{K-LTL}) in the
form
\[\frac{T}{T_F}=\frac{2}{\pi} \left(\frac{K}{\epsilon_F}-0.6\right)^{1/2} \]
and replacing $T/T_F$ into Eq. (\ref{IE-FG}). The expression connecting
$S_{thermal}$ and $K$ is
\begin{equation}
S_{thermal}=s \left(\frac{K}{\epsilon_F}-0.6 \right)^{\lambda}
\label{Sther-K}
\end{equation}
with
\[s=\frac{2a}{\pi}=1.6212, \qquad \lambda=0.5 \]

Expression (\ref{Sther-K}) is the same with the corresponding expression
of $S_{cor}(K)$ which is given by Eq. (\ref{Scor-KE-EG}).
The values of the parameter $s$ and $\lambda$ ($s=2.0786$ and
$\lambda=0.6601$ ) of Eq. (\ref{Scor-KE-EG}) are  close to the constants
$s=1.6212$ and $\lambda=0.5$ of Eq. (\ref{Sther-K}). For that reason we
should expect that the same values of $K$ corresponding
either to the temperature
or to the electron correlation lead to similar values for the two entropies
$S_{thermal}$ and $S_{cor}$. The calculated values of $S_{thermal}$,
for the uncorrelated Fermi gas, versus $K$ are shown in Fig. 3. From that
figure, it is seen that for the same values of $K$, $S_{thermal}$ and
$S_{cor}$ take similar values, as expected.
From the above analysis it is seen that the information entropy sum and the
thermal part of it are related with fundamental quantities, such as,
the temperature, the
thermodynamical entropy and the mean kinetic energy of the system.

\subsection{Thermal effects in electron gas for $T \gg T_F $}
A relation can also be established between the IE and the
thermodynamical entropy in the classical case when it is assumed that
$n(k) \ll 1$. That condition is valid when the density is low
and/or the temperature is high. In that case the MD has the  gaussian form
\cite{Mandl}
\begin{equation}
n(k)=\left(\frac{a}{\pi}\right)^{3/2} e^{-ak^2},\qquad
a=\frac{\hbar^2}{2 m k_B T}
\label{MD-idealgas}
\end{equation}
and is normalized as $\int n(k) d {\bf k}=1$.
The thermodynamical entropy of the system is given by the relation
\cite{Goodstein,Mandl}
\begin{equation}
\frac{S_{TE}}{Nk_B}=\ln V- \ln N + \frac{5}{2}
+\frac{3}{2} \ln \frac{mk_BT}{2 \pi \hbar^2}
\label{TE-IG}
\end{equation}
Following the steps of Section 2, the information entropy sum for the
above system is written
\begin{equation}
S_{IE}=\ln V+ \frac{3}{2}+3 \ln 2 \pi
+\frac{3}{2} \ln \frac{mk_BT}{2 \pi \hbar^2}
\label{IE-IG}
\end{equation}

Comparing Eqs. (\ref{IE-IG}) and (\ref{TE-IG})
and using Eq.(\ref{IE-S0}) a relation between
$S_{TE}$ and $S_{IE}$ can also be found in the case $T \gg T_F$.
That relation has the form
\begin{eqnarray}
S_{IE}&=&S_0+\left(\ln 2-1\right)+\frac{S_{TE}}{Nk_B}
\nonumber\\
&=&\ln N + \left(3\ln 2\pi- 1 \right)+\frac{S_{TE}}{Nk_B}
\label{IE-TE-IG}
\end{eqnarray}
while the thermal part of the information entropy depends on $S_{TE}$
through the relation
\begin{equation}
S_{thermal}=\left(\ln 2- 1\right)+\frac{S_{TE}}{Nk_B}
\label{IE-Thermal-IG}
\end{equation}
Thus the information entropy sum as well the thermal part of it,
in the limit $T \gg T_F $ depends also on the number of electrons
as well as on the thernodynamical entropy of the system. Those
relations are similar to the ones which have been found in the
limit case $T \ll T_F $, only the two constants are different.

From Eqs. (\ref{IE-Thermal-IG}) and  (\ref{TE-IG}) a relation connecting
$S_{thermal}$ with the temperature can be found. That relation has the form
\begin{equation}
S_{thermal}=\frac{3}{2}+\ln \frac{3 \pi^{1/2}}{4}
+\frac{3}{2} \ln \frac{T}{T_F}
\label{Stherm-T}
\end{equation}
Finally, from the well known result
\begin{equation}
K=\frac{\hbar^2}{2m} \int n(k) k^2 d{\bf k}=\frac{3}{2}k_BT
\label{K-idealgas}
\end{equation}
and Eq. (\ref{Stherm-T}) a relation connecting $S_{thermal}$ and
$K$ can be found. That relation has the form
\begin{equation}
S_{thermal}=\frac{3}{2}+\frac{1}{2}\ln \frac{\pi}{6}
+\frac{3}{2} \ln \left(\frac{K}{\epsilon_F}\right)
\simeq 1.1765+\frac{3}{2} \ln \left(\frac{K}{\epsilon_F}\right)
\end{equation}

We can conclude that, at the classical limit, the IE as well as
its thermal part is related to $S_{TE}$, $T$ and $K$, as
in the low temperature limit.

\section{Concluding remarks and Summary}
A study of Shannon's information entropies in position ($S_r$) and
momentum ($S_k$) spaces for three correlated Fermi systems i.e.
nuclear matter, electron gas and liquid $^3$He, was made. The
analysis was performed applying the LOA for the calculation of the
MD in nuclear matter, and using the results of Refs. \cite{Paola-02} and
\cite{Moroni97} for the electron gas and liquid $^3$He, respectively.
 The strength of the fermion correlations in nuclear matter is
measured by the wound parameter, $k_{dir}$, while for the electron
gas and liquid $^3$He by the value of the constant density of the uniform
systems. That strength can be measured also in the same way, in
the three systems,  by the discontinuity gap, $Z_F$ (or 1$-Z_F$),
of the MD at $k=k_F$.

It was found that the information entropy sum, $S=S_r + S_k$,
depends linearly on the logarithm of the number of fermions. There
is also a dependence of $S$ on the strength of correlations. It is
remarkable that for the three different Fermi systems with
different particle interactions, the same or similar two
parameters formulae exist connecting the correlated part,
$S_{cor}$, of the IE of the system with the various kind of the
parameters of the system which measure the strength of the
interactions. For nuclear matter, electron gas and liquid $^3$He
the corresponding expressions are: $S_{cor}=sk_{dir}^{\lambda}$,
$S_{cor}=sr_{s}^{\lambda}$, and $S_{cor}=s\rho^{\lambda}$.
Our results for electron gas are in agreement with the ones of
Ziesche \cite{Ziesche95}. For nuclear matter and electron gas the
dependence of $S_{cor}$ on $Z_F$ is of the form
$S_{cor}=s(1-Z_F)^{\lambda}$, while for liquid $^3$He is of the
form $S_{cor}=s(1-Z_{F}^{\lambda})$. The difference in that
expression of $S_{cor}$  comes from the strong character of
the particle interaction in liquid $^3$He. From the above
dependence of $S_{cor}$ on the various parameters it should become
clear that the values of the IE could be used as a common measure
of the particle correlations of Fermi systems. This is also
supported by the fact that there is the same formula, in the three
systems, which relates $S_{cor}$ with the mean kinetic energy of
the system of the form $S_{cor}=s\left( K/ \epsilon_F -
0.6\right)^{\lambda}$. The increase of the IE (through $S_{cor}$)
with the parameter $1-Z_F$ or with the other correlation
parameters ($k_{dir}$ or $\rho$ or $r_s$) is due to the fact that
the effect of the particle correlations is to diffuse the MD from the
step functional form (ideal fermi gas) creating a long tail
behaviour for $k>k_F$. That diffusion of the MD leads to a decrease of
the order of the system (in comparison to the order step function
MD), thus, it leads to an increase of the information content of
the system.

We studied, also, how the thermal effects affect the information
entropy sum of the uncorrelated electron gas. The study was made
for two cases, the low temperature limit and the high one. It was
found that, in both cases, there are relations which connect the
thermal part of the IE with the fundamental quantities such as the
temperature, the thermodynamical entropy and the mean kinetic
energy. The dependence of $S_{thermal}$ on $T$ and $S_{TE}$ is
linear with a larger slope in the low temperature limit than in the
high one. $S_{thermal}$ depends on the logarithm of $K$ in the
high temperature limit, while in the low one is of the form
$S_{thermal}=(2\alpha/\pi)\left( K/ \epsilon_F -
0.6\right)^{\lambda}$, where $\alpha$ is the slope of the linear
expression which relates $S_{thermal}$ with $T$. That expression
is almost the same with the one which holds for the correlated
electron gas, $S_{cor}=s\left( K/ \epsilon_F -
0.6\right)^{\lambda}$. The values of the parameters $s$ and
$\lambda$ are very close to the constants $2\alpha/\pi$ and 0.5,
respectively. Thus, independent of the reason that causes the
increase of $K$ the information entropy increases almost by the
same amount either in the correlated electron gas or in the
uncorrelated one.

\begin{center}
{\bf ACKNOWLEDGMENTS}
\end{center}
The authors would like to thank Dr. Paola Gori-Giorgi for
providing the data for the correlated electron gas and Dr. Saverio
Moroni for providing the data for the liquid $^3$He. They would
like also to thank Dr. C.P. Panos for fruitful discussions.


\newpage
\begin{figure}
\begin{center}
\begin{tabular}{ccc}
{\epsfig{figure=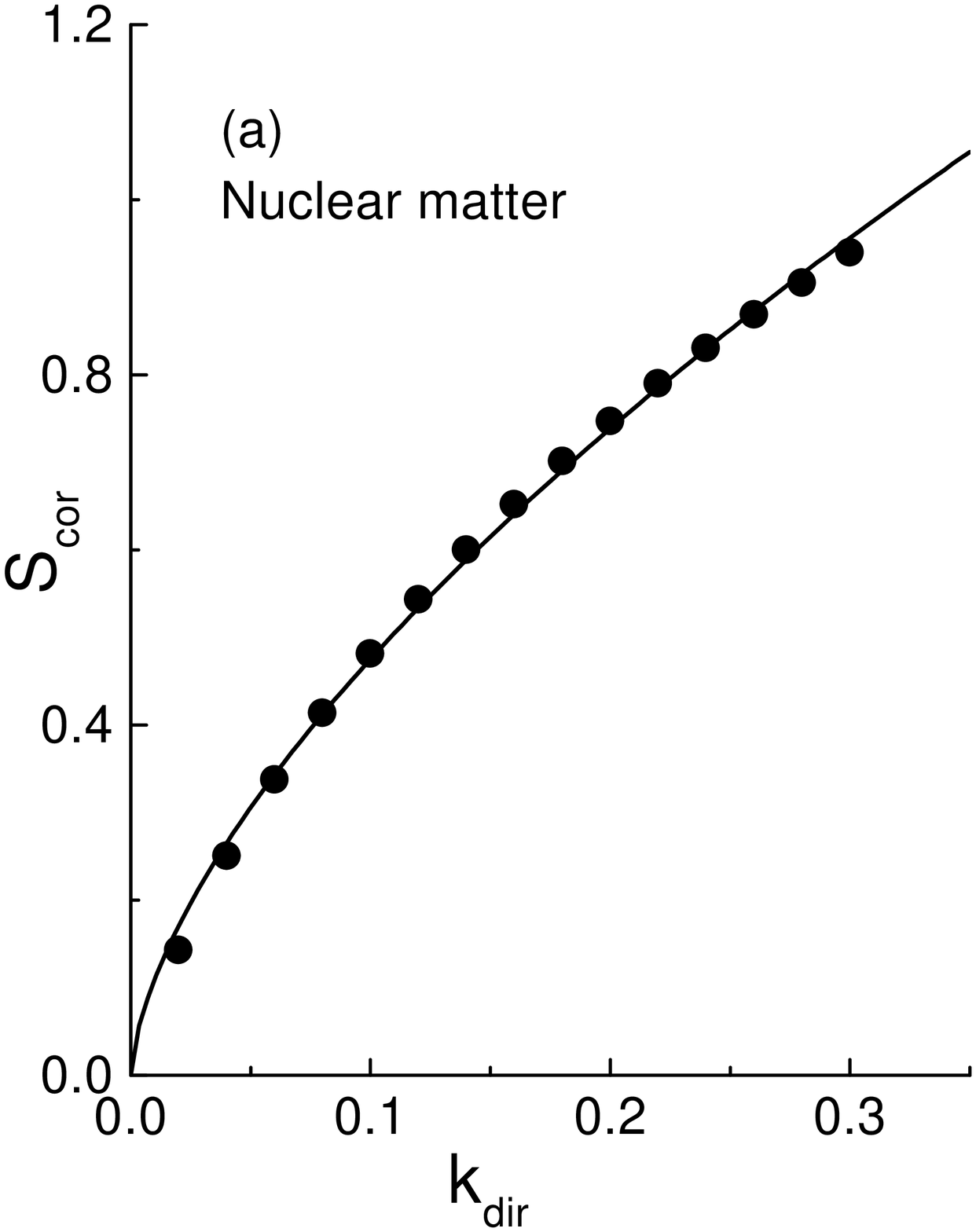,width=5.5cm} }&
{\epsfig{figure=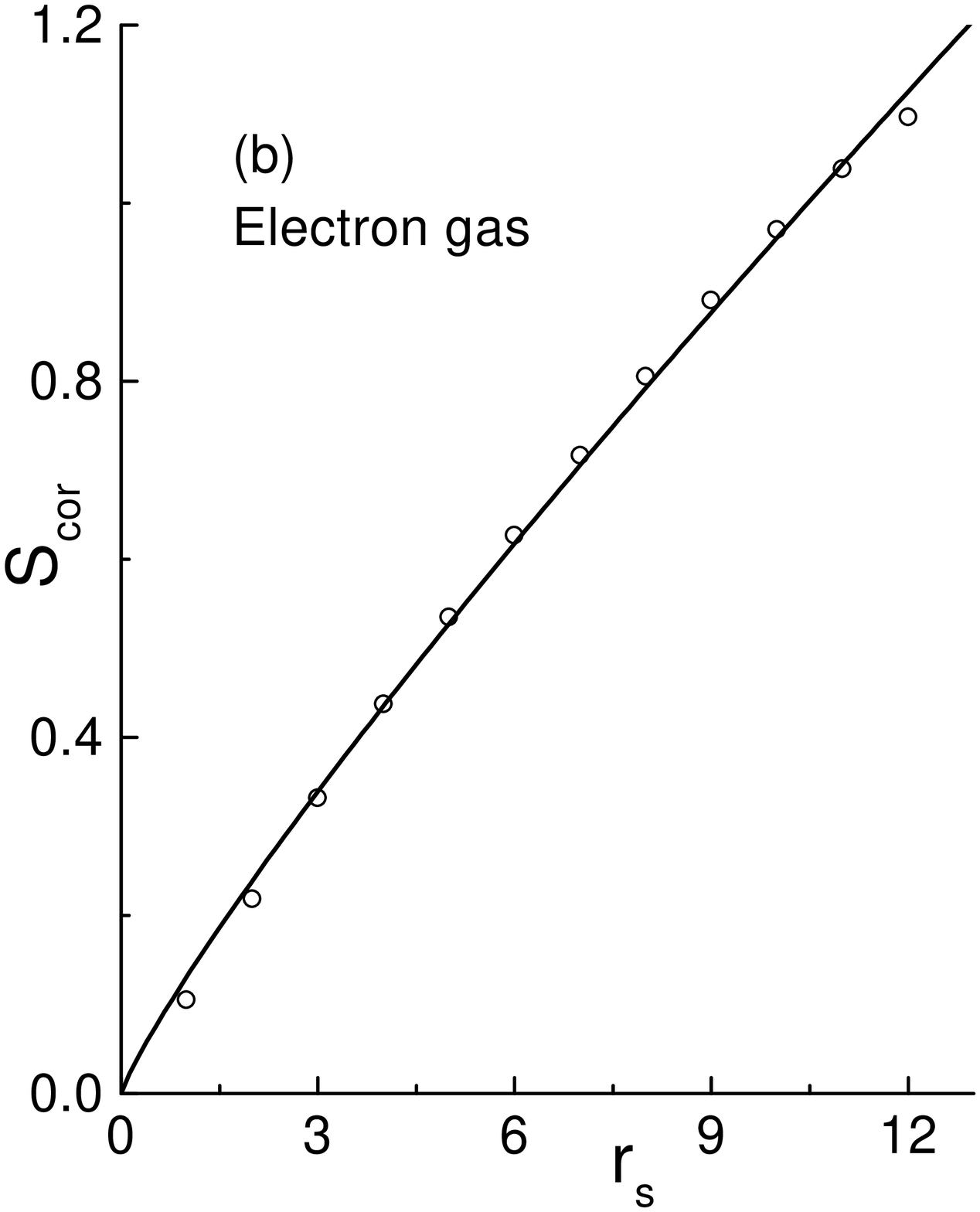,width=5.5cm} }&
{\epsfig{figure=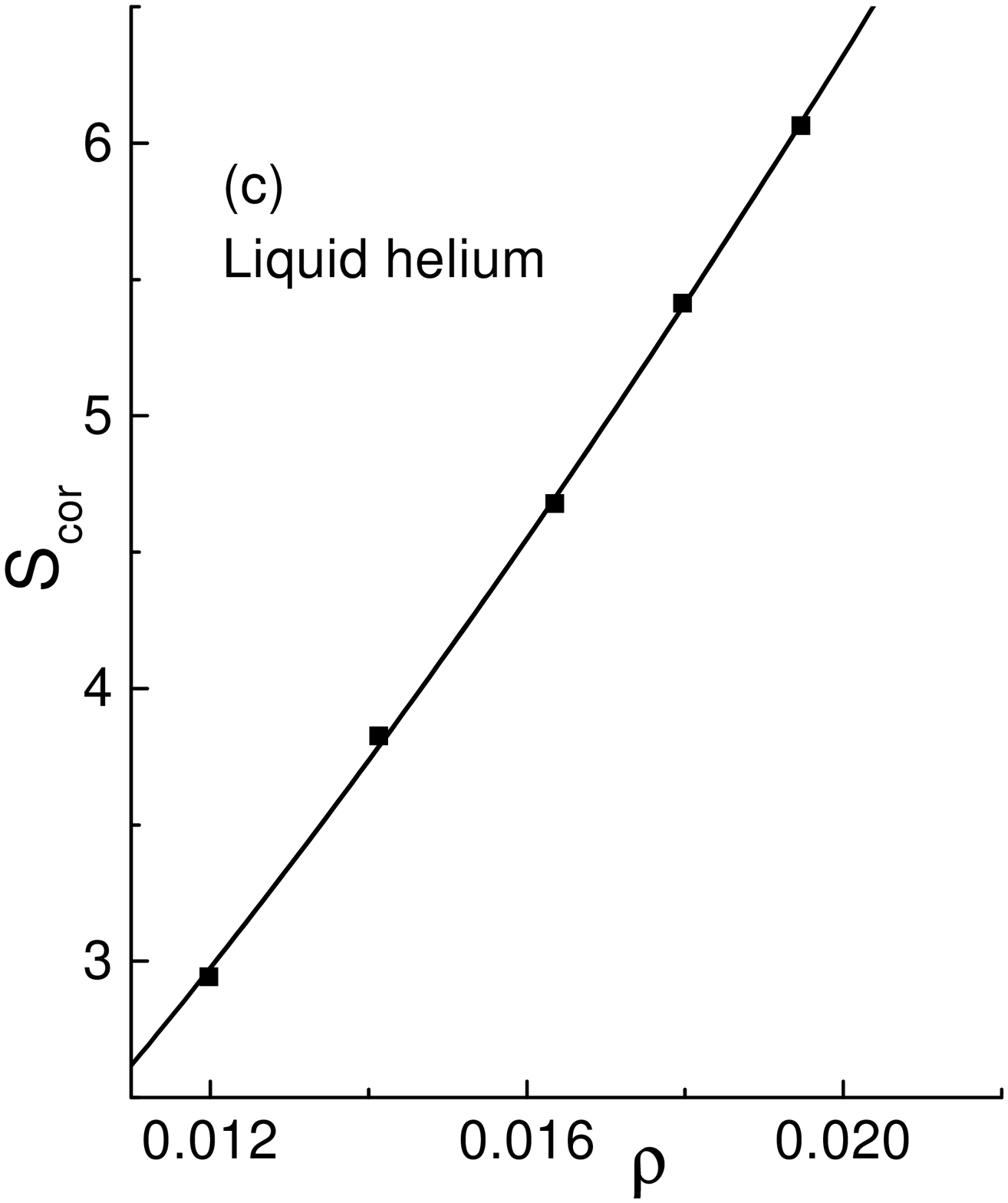,width=5.5cm} }
\end{tabular}
 \vspace*{-5mm}
\caption{The correlated part of the information entropy, $S_{cor}$,
for a) nuclear matter, b) electron gas, and c) liquid $^3$He versus
the wound parameter $k_{dir}$, the effective radius $r_s$,
and the density $\rho$, respectively.}
\end{center}
\end{figure}
\begin{figure}
\begin{center}
\begin{tabular}{c}
\hspace*{-1.5cm}{\epsfig{figure=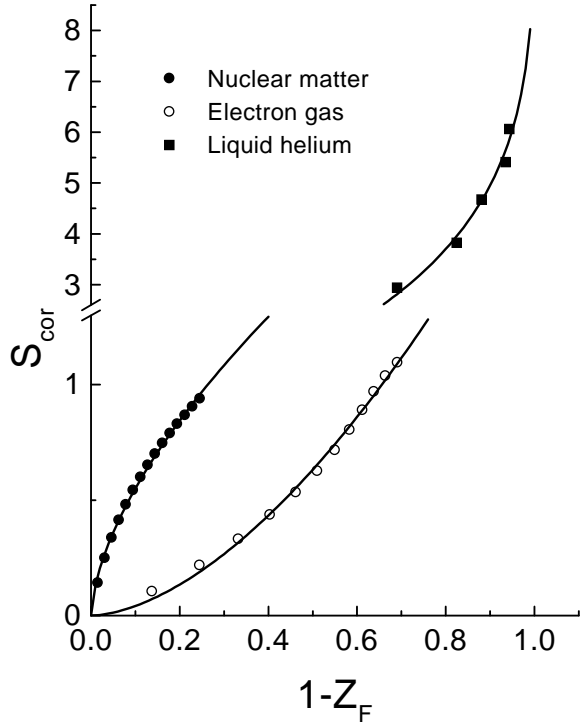,width=8.5cm} }
\end{tabular}
 \vspace*{-5mm}
\caption{The correlated part of the information entropy for
various Fermi systems versus the discontinuity parameter $1-Z_F$.}
\end{center}
\end{figure}
\begin{figure}
\begin{center}
\begin{tabular}{c}
\hspace*{-1.5cm}{\epsfig{figure=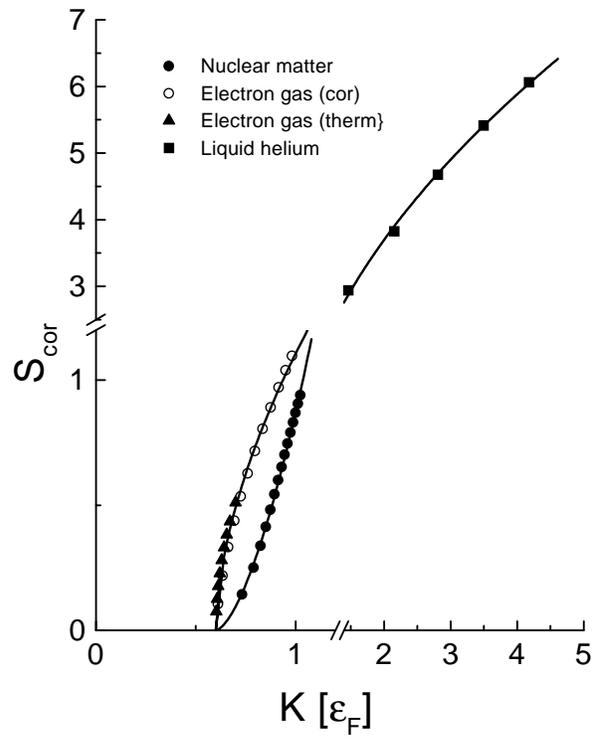,width=8.5cm} }
\end{tabular}
 \vspace*{-5mm}
\caption{The correlated part of the information entropy for
various Fermi systems and its thermal part for electron gas
versus the mean kinetic energy in units of Fermi energy.}
\end{center}
\end{figure}
\begin{figure}
\begin{center}
\begin{tabular}{cc}
\hspace*{-1.5cm}{\epsfig{figure=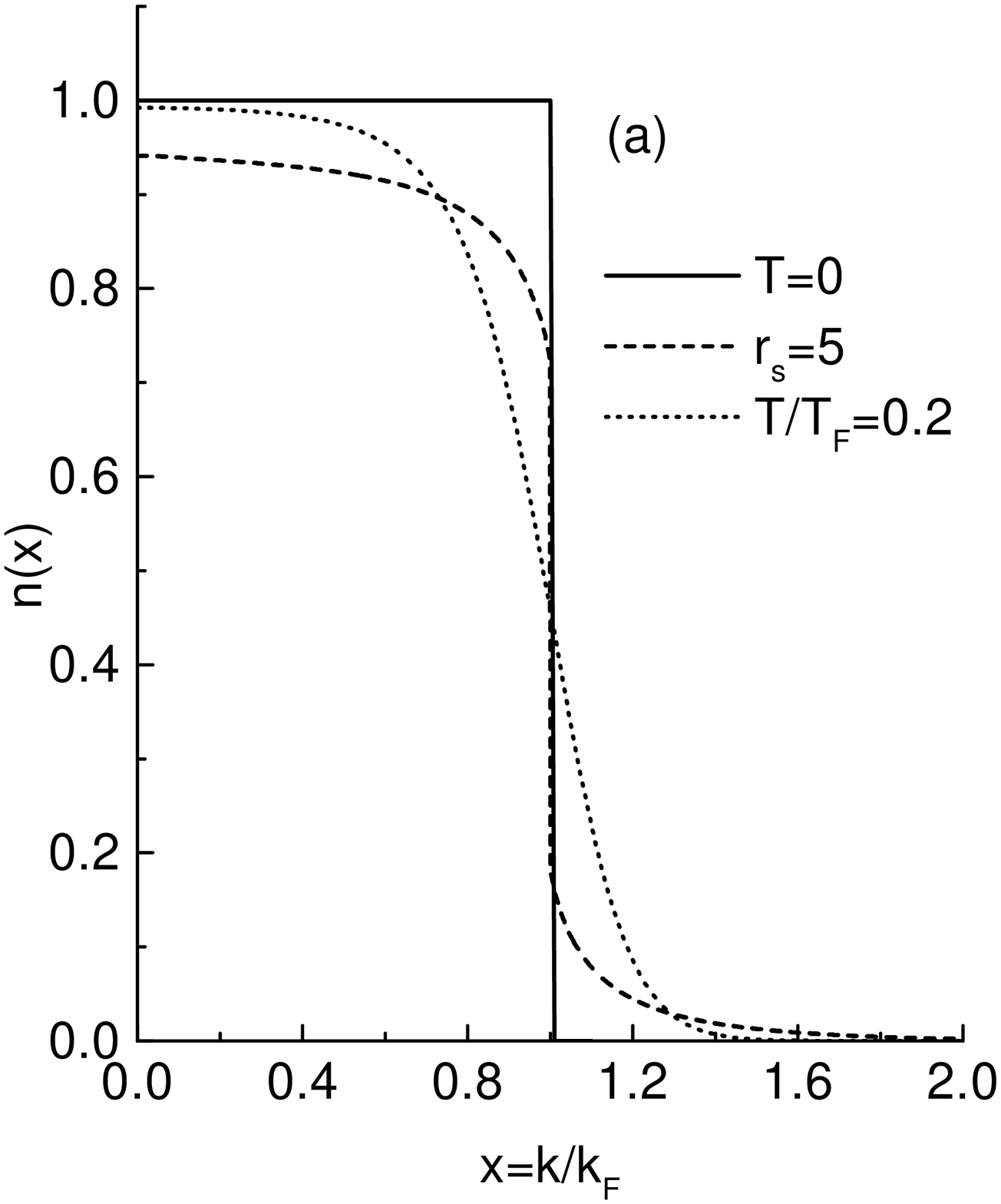,width=6.5cm} }&
{\epsfig{figure=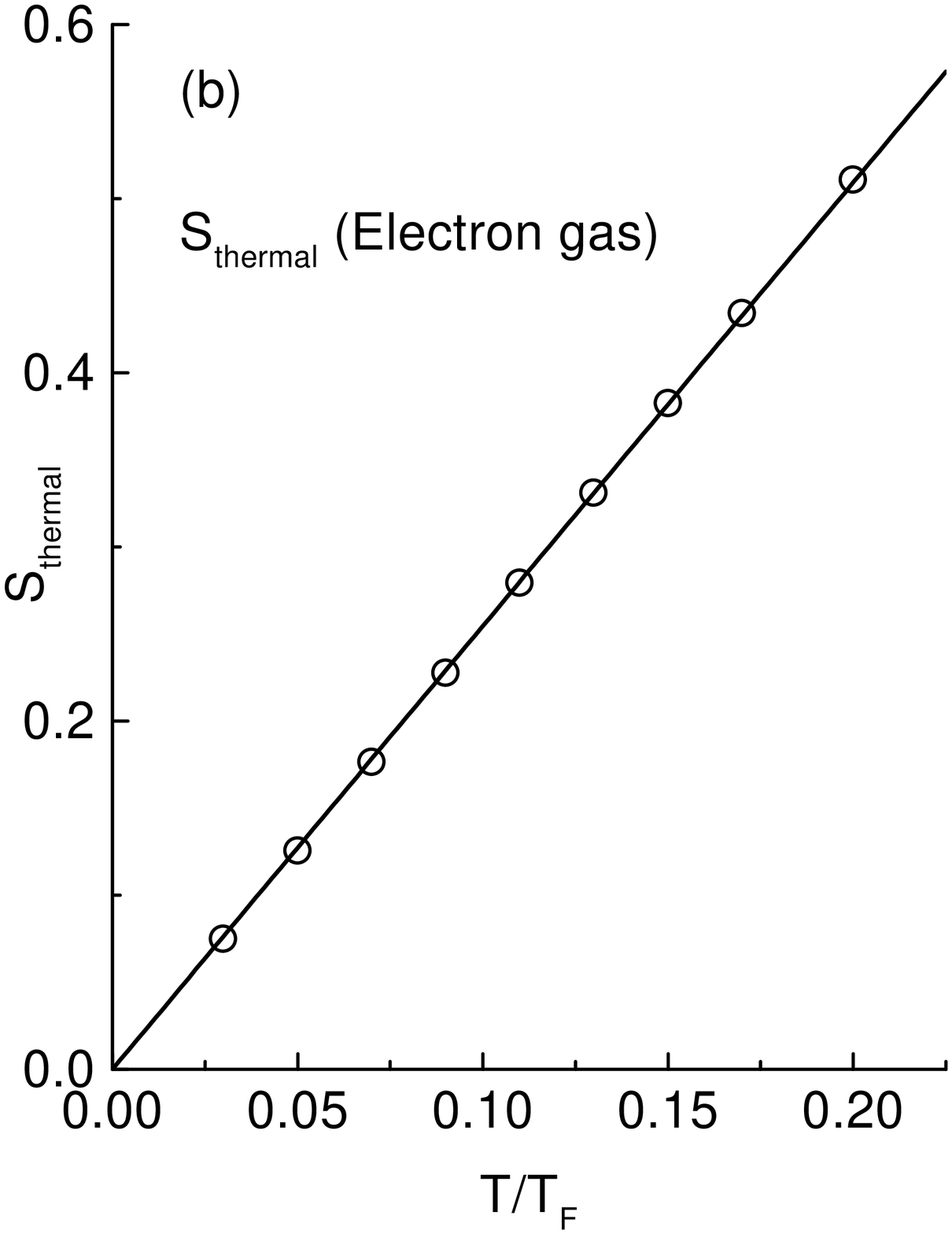,width=6.5cm} }
\end{tabular}
 \vspace*{-5mm}
\caption{ a) The momentum distribution for correlated electron gas with
effective radius $r_s=5$ and the uncorrelated one for temperature $T=0$
and $T=0.2 T_F$ versus the ratio $x=k/k_F$.
b) The thermal part of the information entropy versus the temperature $T$
in units of $T_F$.}

\end{center}
\end{figure}


\end{document}